\documentclass[aip,reprint,12pt,onecolumn,a4paper,longbibliography,groupedaddress,tightenlines,floatfix]{revtex4-2}

\usepackage{graphicx}
\usepackage{rotating}
\usepackage{microtype}
\usepackage{amsmath, amssymb}
\AtBeginDocument{\usepackage{booktabs}}

\usepackage[T1]{fontenc}

\usepackage[scale=0.9]{tgheros}\usepackage{tgtermes}\usepackage{newtxmath}\usepackage[scale=0.94]{inconsolata}
\usepackage[siunitx,american]{circuitikz}
\usetikzlibrary{calc,arrows.meta}

\usepackage{xcolor}
\definecolor{linkblue}{RGB}{0,90,100}
\usepackage[colorlinks=true, citecolor=linkblue, urlcolor=linkblue, linkcolor=black]{hyperref}
\hypersetup{pdftitle={Bayesian inference with a polymer probabilistic computer}}
\usepackage{cleveref}\crefname{equation}{Eq.}{Eqs.}

\newcommand{\kl}[2]{D_{\mathrm{KL}}\mathopen{}\left(#1 \,\|\,  #2\right)\mathclose{}}

\newcommand{\Vmvd}{V_{\mathrm{mvd}}}
\newcommand{\Vhat}{\widehat V_{\mathrm{mvd}}}
\newcommand{\Vref}{V_{\mathrm{ref}}}
\newcommand{\Voff}{V_{\mathrm{offset}}}
\newcommand{\VDD}{V_{\mathrm{DD}}}
\newcommand{\Vc}{V^\circ}

\begin{document}

\title{Bayesian inference with a polymer probabilistic computer}


\author{Travis Wanless}
\affiliation{Center for Optical Materials Science and Engineering Technologies (COMSET), Clemson University, Clemson, SC 29634, USA}
\affiliation{Department of Materials Science and Engineering, Clemson University, Clemson, SC 29634, USA}

\author{Yuriy Bandera}
\affiliation{Center for Optical Materials Science and Engineering Technologies (COMSET), Clemson University, Clemson, SC 29634, USA}
\affiliation{Department of Materials Science and Engineering, Clemson University, Clemson, SC 29634, USA}

\author{Lubomir Kostal}
\email[Secondary corresponding author: ]{lubomir.kostal@fgu.cas.cz}
\affiliation{Institute of Physiology, Computational Neuroscience Group, Czech Academy of Sciences, 142 20 Prague 4, Czech Republic}

\author{Stephen H. Foulger}
\email[Primary corresponding author: ]{foulger@clemson.edu}
\affiliation{Center for Optical Materials Science and Engineering Technologies (COMSET), Clemson University, Clemson, SC 29634, USA}
\affiliation{Department of Materials Science and Engineering, Clemson University, Clemson, SC 29634, USA}
\affiliation{Department of Bioengineering, Clemson University, Clemson, SC 29634, USA}

\begin{abstract}
Bayesian inference weighs competing explanations of noisy data by sampling from a posterior distribution, a task that deterministic hardware performs only by simulating randomness. 
Probabilistic bits (p-bits) supply this randomness physically, but the few hardware networks of p-bits built so far use inorganic devices, address optimization or learning, and are judged without reference to the spread expected from finite sampling. Here we report a probabilistic computer built from solution-processed conjugated-polymer memristors. Four polymer p-bits, tracked online for drift and coupled in a closed Gibbs-sampling loop, sample competing, degenerate and cooperative posteriors over sixteen states with statistics that fall within the range of ideal p-bits, both in their distribution and in their autocorrelation, without any parameter fitted to the target. This agreement extends to new simulated sensor readings whose outcome was predicted before each run, including readings that point away from the true answer. 
Reprogrammed in software, the same circuit and controller reproduces the invertible AND gate of earlier p-bit networks, with its temperature fixed in advance rather than fitted. Polymer p-bits thus open a route to printed, large-area probabilistic computers that infer alongside their sensors.
\end{abstract}

\maketitle

The Boltzmann machine is a model of collective computation in neural networks, consisting of stochastic binary units whose joint states follow a Boltzmann distribution\cite{r:hopfield82,r:ackley85}. 
Sampling from this distribution enables probabilistic inference by allowing competing explanations of noisy data to be weighed against one another\cite{r:geman84,r:mackay-book}.

Probabilistic bits (p-bits) physically realize these stochastic units\cite{r:camsari17prx,r:camsari19,r:kaiser21,r:chowdhury23}. Several  hardware networks of physical p-bits have been reported. Eight stochastic magnetic tunnel junctions (MTJs), coupled through a microcontroller, factorized integers up to 945\cite{r:borders19}. Five such p-bits, with weights stored on capacitors, were trained in situ to reproduce the truth table of a full adder\cite{r:kaiser22}. Four HfO$_2$ ferroelectric tunnel junctions, coupled through a ferroelectric crossbar, solved a maximum-independent-set problem on a four-node graph\cite{r:guan26}, and three volatile Ag/SiO$_2$ memristors, coupled through a crossbar of non-volatile memristors, reproduced invertible AND and NAND gates\cite{r:soh26}. Three MTJ p-bits coupled directly through a resistor network reproduced an invertible AND gate\cite{r:lv19}, and 80 superparamagnetic MTJs coupled through a microcontroller solved traveling-salesman problems\cite{r:si24}. Voltage-controlled MTJs integrated into a 96-kb CMOS memory chip solved Max-cut and circuit-routing problems by annealing\cite{r:li26ne}, and 250 MTJs   served as random sources for factorization and Max-cut on a field-programmable gate array\cite{r:yang26}.
These demonstrations focused on optimization or on learning a fixed mapping. Their sampled distributions were assessed using histograms compared with simulations that had one fitted parameter\cite{r:borders19}, a Kullback--Leibler (KL) divergence from the target distribution, without the range expected from finite sampling\cite{r:kaiser22}, the predominance of correct states\cite{r:guan26, r:soh26}, the optimality gap of the final state\cite{r:li26ne},  or the success probability over repeated annealing runs\cite{r:yang26}. Other stochastic hardware has supported probabilistic computation without forming networks of physical p-bits: stochastic MTJs have clocked digital p-bits on a field-programmable gate array\cite{r:singh24}, and voltage-controlled MTJs have supplied the random bits of digital p-bits on a CMOS chip\cite{r:duffee25}; memristor arrays have supplied the random variables of a Markov chain Monte Carlo algorithm through their cycle-to-cycle conductance variability\cite{r:dalgaty21}, stored the likelihoods of a Bayesian machine\cite{r:harabi22} or the couplings of a CMOS probabilistic computer\cite{r:he26}, and provided the annealing noise of a Hopfield network\cite{r:cai20}; and ferroelectric transistors have served as stochastic synapses of a neural sampling machine\cite{r:dutta22}.

Many materials have now been shown to act as p-bits, but almost all have been tested only as single devices whose measured response was then used in simulated networks. Such p-bits with a tunable sigmoidal response include diffusive memristors\cite{r:woo22,r:baek24}, MTJs integrated with 2D transistors\cite{r:daniel24}, perovskite nickelates\cite{r:park22}, NbO$_x$ Mott oscillators\cite{r:rhee23}, silicon biristors\cite{r:kim24}, manganite nanowires\cite{r:wang25} and organic--inorganic photonic heterojunctions\cite{r:seo26}. Ferroelectric field-effect transistors have been proposed as p-bits on the basis of phase-field simulations\cite{r:luo23}. Networks of physically coupled p-bits remain rare, and the few reported to date are all based on inorganic devices\cite{r:borders19,r:lv19,r:kaiser22,r:si24,r:guan26,r:li26ne,r:soh26,r:yang26}.

Among soft materials, polymeric memristors built from conjugated backbones carrying electroactive pendant groups have been developed over the last several years. 
In these devices, conduction through the backbone and through the pendant groups forms coupled percolation networks that make the response depend on voltage history\cite{r:foulger21jmcc}. 
Varying the backbone and pendant chemistry within this design, such devices have been used to generate random bit streams\cite{r:foulger23ted} and, more recently, to realize single p-bits\cite{r:foulger25apr}, 
whose stochastic response has been traced to the polymer's molecular structure\cite{r:foukos26apr}. 
Related polymer devices have also been engineered at the molecular level\cite{r:vilkos26ceja}, modeled\cite{r:koshab26nce} and studied toward hardware spiking neural networks\cite{r:foukos26acs}.

\begin{figure*}
\makebox[\textwidth][c]{\includegraphics{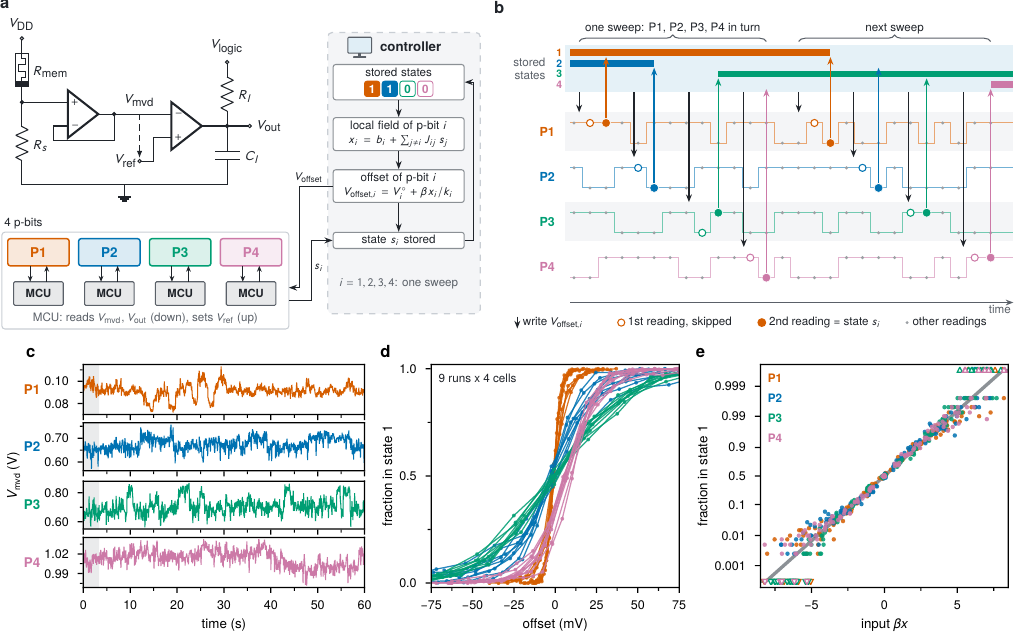}}
\caption{\textbf{The polymer p-bit computer
with four p-bit primitives and its measured response.}
\textbf{a}, Polymer p-bit primitive. The memristor $R_{\mathrm{mem}}$
and load resistor $R_s$ form a voltage divider; the buffered midpoint
voltage $V_{\mathrm{mvd}}$ is compared with a reference
$V_{\mathrm{ref}}$, which is periodically updated from
$V_{\mathrm{mvd}}$ (dashed), to produce the binary output
$V_{\mathrm{out}}$. The p-bit is in state 1 while $V_{\mathrm{mvd}}$
lies below $V_{\mathrm{ref}}=\widehat
V_{\mathrm{mvd}}+V_{\mathrm{offset}}$. A microcontroller (MCU) per p-bit
reads $V_{\mathrm{mvd}}$ and $V_{\mathrm{out}}$, keeps the running
baseline $\widehat V_{\mathrm{mvd}}$ and sets $V_{\mathrm{ref}}$; it
receives the offset from the controller and returns the readings.  The
unit has four such p-bits (P1--P4 at 
$V_{\mathrm{DD}}$ of 2.0, 2.0, 1.5
and 2.0\,V). The controller keeps the stored states and updates the
p-bits in turn, one sweep at a time: for p-bit $i$ it computes the local
field $x_i$ from the other stored states, applies the offset
$V_{\mathrm{offset},i}$  and
stores the read state as $s_i$.
\textbf{b}, The update cycle: the controller writes the offset
of p-bit $i$ (arrow down); the comparators read all the time. The first
reading after the write is skipped (open circle) and the second is
stored as $s_i$ (filled), which replaces the stored state (arrow up);
then the next p-bit, four updates per sweep. The traces are
illustrative.
\textbf{c}, Raw $V_{\mathrm{mvd}}$ of the four p-bits over an
illustrative minute of the first V2 run without level changes (gray:
four sweeps), one reading every 63\,ms.  \textbf{d}, Fraction in state 1 against the
offset in the fine calibration (400 readings per point) of
each of the nine runs (Methods).  
\textbf{e}, The same points as in \textbf{d}, plotted against the input $\beta x$ obtained from the logistic fit of each p-bit in each run, on a logit axis. Triangles mark offsets at which all readings were in state 0 or all in state 1, which cannot be shown on a logit axis. Gray line: ideal p-bit response, $1/(1+e^{-\beta x})$.} \label{fig:unit}
\end{figure*}

Here we report a probabilistic computer built from these polymer p-bits, to our knowledge the first network of p-bits made from soft matter, and apply it to Bayesian posterior inference\cite{r:mackay-book} rather than optimization. 
Since p-bits, like Boltzmann machines, are inspired by models of neural computation,
we test the computer on a task that nervous systems solve. We consider a chemical-mixture identification problem modeled on olfaction, in which the computer infers which of four chemicals are present from a single noisy reading of four cross-reactive sensors, as the olfactory system must do with its receptors\cite{r:persaud82,r:grabska17}, and returns the full posterior over all possible mixtures rather than a single best answer.
Four polymer p-bits, coupled in a closed hardware sampling loop, sample competing, degenerate and cooperative posteriors over sixteen states.
In each case the KL divergence\cite{r:cover} of the output distribution to the exact posterior falls within the finite-sample envelope of simulated ideal samplers, and the autocorrelation of the sampled states follows that of ideal Gibbs sampling\cite{r:geman84}. Because each p-bit is defined by a solution-processed polymer film rather than a lithographically patterned junction, the approach is not confined to silicon fabrication and points toward larger, printed probabilistic arrays that could compute alongside the sensors whose data they interpret. 
Fig.~\ref{fig:unit} presents
a schematic of the computer and the response of its p-bits.

\subsection*{Hardware implementation of the polymer p-bit computer}

Fig.~\ref{fig:unit}a presents the architecture of the polymer p-bit computer. 
The current implementation comprises four p-bit primitives, whose circuit has been described previously\cite{r:foukos26apr}. Briefly, each primitive contains a polymer memristor $R_{\mathrm{mem}}$ in the upper arm of a voltage divider, where the active layer is pB6CC, a conjugated polymer reported previously\cite{r:foukos26acs} (Supplementary Fig.~\ref{fig:Spolymer}).
The buffered midpoint voltage $V_{\mathrm{mvd}}$ is compared with a reference $V_{\mathrm{ref}}$, and the comparator yields a binary state $s_i$. 
A microcontroller attached to each p-bit reads $V_{\mathrm{mvd}}$ and $V_{\mathrm{out}}$ and sets $V_{\mathrm{ref}}$. 
A central controller holds the four stored states, computes the local field $x_i=b_i+\sum_{j\neq i}J_{ij}s_j$ of the p-bit being updated, and converts it into a comparator offset 
\begin{equation}
V_{\mathrm{offset},i}=V_i^{\circ}+\frac{\beta x_i}{k_i}.
\label{eq:offset}
\end{equation}
Here $b_i$ is the bias of p-bit $i$ and $J_{ij}$ its coupling to p-bit $j$, both set by the task, and $\beta$ is the inverse temperature parameter. The midpoint $V_i^{\circ}$ is the tracked offset at which p-bit $i$ is equally likely to be 0 or 1, and $k_i$ is the tracked slope of its logistic response (Methods).
The local field $x_i$, which combines the data-dependent bias and the coupling terms, is applied solely by shifting the comparator reference. The memristor itself is never driven by the input; its supply $V_{\mathrm{DD}}$ is held fixed throughout sampling.
The p-bits are updated sequentially, one sweep comprising one update of each p-bit in turn (Fig.~\ref{fig:unit}b). 
After a new offset is written, the first comparator reading is discarded to let the circuit settle, and the second is stored as the p-bit's state; the remaining readings serve the drift trackers.

\subsection*{Polymer memristors as probabilistic bits}

The stochasticity originates in the polymer itself, whose resistance changes randomly in time and makes $V_{\mathrm{mvd}}$ wander continuously (Fig.~\ref{fig:unit}c). 
The four devices operate at very different $V_{\mathrm{mvd}}$ levels, from about 0.1\,V for primitive 1 (P1) to about 1.0\,V for P4, and with different fluctuation amplitudes. 
Accordingly, their raw transfer curves are significantly different. 
For each p-bit, the probability of a high output rises from 0 to 1 as the comparator offset is swept, but the width of this transition ranges from about 10\,mV to 110\,mV, depending on the device and the run (Fig.~\ref{fig:unit}d, 9 runs $\times$ 4 p-bits).

The polymer devices are not stationary noise sources. Their $V_{\mathrm{mvd}}$ readings are correlated in time: between consecutive readings, 63\,ms apart, the deviations of $V_{\mathrm{mvd}}$ from its running mean over one minute have correlation coefficients of 0.35--0.93, depending on the device. Their operating points also drift over minutes to hours. The controller follows these changes with four running estimates per p-bit (Methods). The baseline $\widehat V_{\mathrm{mvd}}$ and the noise width $q$ (Methods) are updated from every reading, with time constants of 1.5 and 300 readings. 
The midpoint $V_i^{\circ}$ and the scale $\kappa_i$ of the slope $k_i=\kappa_i/q$ are updated only from answers whose target probability lies between 1 and 99\,\%, so that no calibration readings interrupt the sampling. None of the four estimates is fitted to the posterior. In addition, the memristors are grounded for 30\,s every 340\,s, which is intended to limit slow drift of the operating point (Methods).

Once each curve is expressed through the tracked parameters as a function of the dimensionless input $\beta x$, all four devices collapse onto the single logistic function $1/(1+e^{-\beta x})$ (Fig.~\ref{fig:unit}e). 
Plotted on logit axes, which expand the tails near 0 and 1, the measured probabilities follow the ideal logistic line from about 0.001 to 0.999. 
This corresponds to inputs up to $|\beta x|\approx 6$ on either side of zero.
Residual scatter appears only at the extremes, where few updates fall and the expected number of rare outcomes approaches one, so the measured probabilities are limited by counting statistics rather than by the devices.
Thus, although the four polymer devices differ in resistance, noise level and drift, each responds to its input exactly as an ideal p-bit would. The network can therefore treat them as identical, which is what allows it to sample the posteriors reported below. 

\subsection*{Odor identification as a Boltzmann sampling problem}

\begin{figure*}
\makebox[\textwidth][c]{\includegraphics{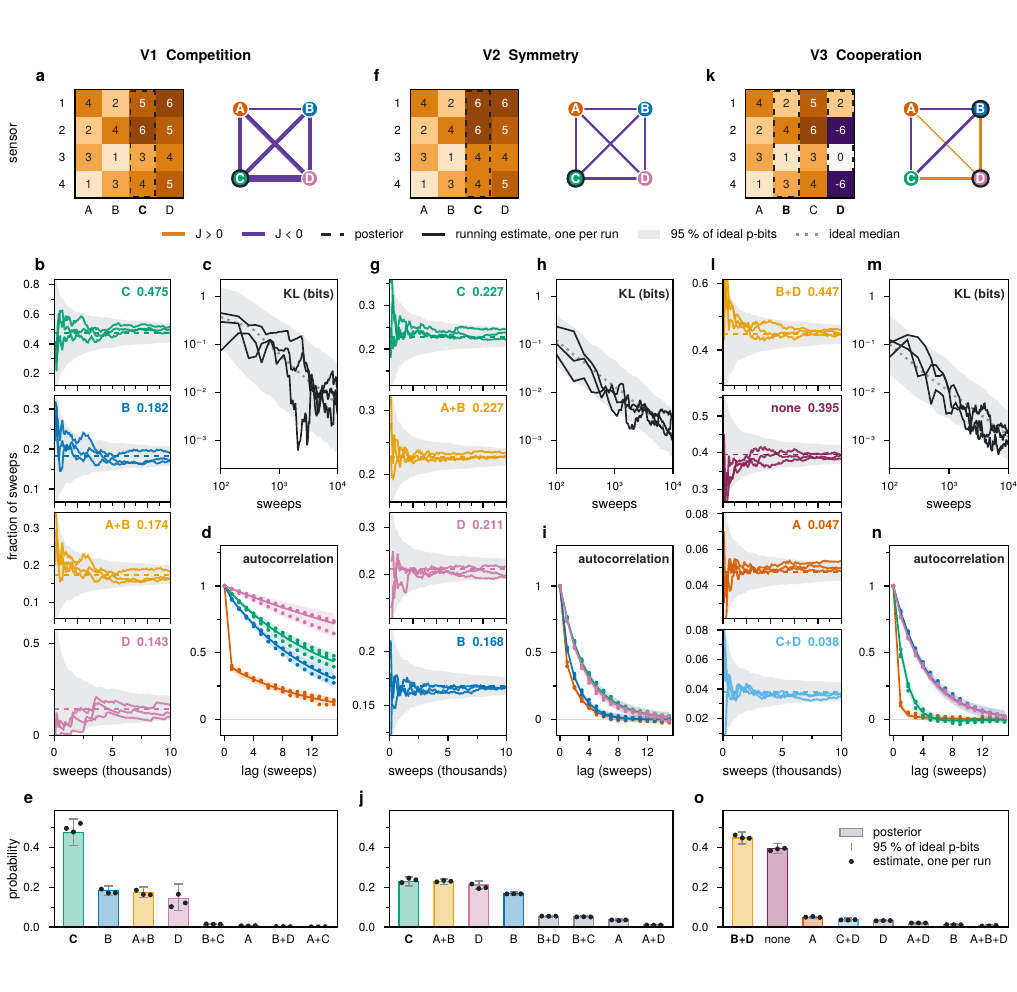}}
\caption{\textbf{The four polymer p-bits sample three posteriors of
different structure.} One column per task: V1 competition
(\textbf{a}--\textbf{e}), V2 symmetry (\textbf{f}--\textbf{j}), V3 cooperation
(\textbf{k}--\textbf{o}); three runs per task on the same reading,
the first 10\,000 sweeps of
each.  The chemicals A--D of each task (see Methods section
`Inference tasks') are assigned to p-bits 1--4 of Fig.~\ref{fig:unit}.
\textbf{a,f,k}, Sensor matrix $F$ (dashed and bold: the chemicals present) and
the couplings $J$ of the Boltzmann form of the posterior (line width
proportional to $|J|$; orange positive, purple negative; ring: chemical
present).  \textbf{b,g,l}, The fraction of each of the four most
probable joint states over the sweeps (one running estimate per run, its
posterior dashed, gray the range of 95\,\% of runs of ideal p-bits).
\textbf{c,h,m}, The Kullback--Leibler divergence from the posterior,
$\sum_sv(s)\log_2[v(s)/P(s)]$ over visited joint states, against the
number of sweeps, with the median (dotted) and 95\,\% range of ideal
runs, on one scale for the three tasks.  \textbf{d,i,n}, The
autocorrelation of the state of each p-bit between sweeps (dots, one set
per run; the colors of the p-bits in \textbf{a,f,k}), that of an ideal
Gibbs sampler, computed exactly (line; Supplementary Note~3), and the range of 95\,\% of ideal
runs (band).  \textbf{e,j,o}, Posterior of the eight most probable
joint states (bars; the four of \textbf{b,g,l} in their colors), the
range into which 95\,\% of 1000 simulated runs of ideal p-bits with the
same 10\,000 sweeps fall (whiskers), and the fraction of sweeps of each
run (dots).}
\label{fig:tasks}
\end{figure*}

We tested the computer on three inference tasks whose posteriors differ in structure 
(Fig.~\ref{fig:tasks}).
The tasks, V1 (competition), V2 (symmetry) and V3 (cooperation), are versions of the same odor-identification problem, a form of perceptual inference from overlapping, noisy sensory inputs of the kind that brains perform and that artificial systems must perform at the sensor\cite{r:fiser10,r:grabska17}. They differ only in how the sensors respond to the chemicals, and hence in how alternative explanations of a reading relate to one another. 
Because each task has only sixteen possible answers, its exact posterior can be computed by enumeration, and this exact solution is the reference against which the hardware is judged.
In each task, four sensors respond to chemicals A--D according to a response matrix $F$, and the chemicals present are inferred from a single reading $y=Fs^{*}+\sigma\xi$, 
in which each sensor reading carries a random Gaussian error of standard deviation $\sigma$, independent of the errors of the other sensors.
With all mixtures equally likely a priori, the posterior probability of each of the $2^4$ mixtures $s\in\{0,1\}^4$ ($s_i=1$ if chemical $i$ is present) decreases with the mismatch between the reading and the response that the mixture predicts, $P(s\mid y)\propto\exp\!\big(-\|y-Fs\|^2/2\sigma^2\big)$.
Because $s_i^2=s_i$, it takes the Boltzmann form $P(s\mid y)\propto\exp\!\big(\sum_i b_i s_i+\tfrac{1}{2}\sum_{i\neq j}J_{ij}s_is_j\big)$. The couplings $J_{ij}=-(F^{\top}F)_{ij}/\sigma^2$ depend only on the sensors, and the biases $b_i=(F^{\top}y)_i/\sigma^2-(F^{\top}F)_{ii}/2\sigma^2$ contain the reading, so that a new reading changes only the biases. The overlaps between the columns of $F$ determine the sign and magnitude of the couplings and hence the structure of the posterior. In the competition task (V1, Fig.~\ref{fig:tasks}a), all couplings are negative. 
Chemical C, which is present, has posterior probability 0.475, and the alternatives B, A+B and D have 0.14--0.18 each. 
In the symmetry task 
(V2, Fig.~\ref{fig:tasks}f), the response to C equals the sum of the responses to A and B. C alone and A+B together then explain any reading equally well, and the posterior has two equal maxima (0.227 each) that differ in three bits. 
In the cooperation task 
(V3, Fig.~\ref{fig:tasks}k), the response to D partly cancels the responses to the other chemicals, which makes all couplings of D positive. The posterior is bimodal, with 0.447 on the mixture present, B+D, and 0.395 on the absence of all four chemicals. The three tasks thus test different properties of a sampler. In V1 it must move between explanations that exclude one another, in V2 it must divide its sweeps equally between two explanations that no reading can distinguish, and in V3 it must switch B and D together through intermediate states of low probability.

On the computer, each chemical is assigned to one p-bit, and the biases and couplings enter the local field $x_i$ that sets the comparator offset of that p-bit, so that ideal p-bits would visit each mixture in proportion to its posterior probability. The readings, noise levels and the construction of $b$ and $J$ are given in Methods.

\subsection*{Posterior sampling at ideal-p-bit fidelity}

We compared the hardware with ideal p-bits, which follow 
Eq.~(\ref{eq:conditional}) of the Methods exactly, in three ways: how quickly the sampled distribution approaches the posterior, how closely it matches the posterior after 10\,000 sweeps, and whether successive states are correlated as in an ideal Gibbs sampler.
Each task was run three times on the same reading, that is, with the same draw of the sensor noise $\xi$, which tests whether the result is reproducible from run to run. The KL divergence between the visit frequencies and the exact posterior decreased with the number of sweeps along the median of ideal runs. It lay within their 95\,\% range at every checkpoint, taken every 100 sweeps, in eight of the nine runs, and at 99 of the 100 checkpoints in the ninth (Fig.~\ref{fig:tasks}c,h,m). 
The running estimates of the four states shown lay within their ranges at 83--100\,\% of the checkpoints 
(Fig.~\ref{fig:tasks}b,g,l). 
After 10\,000 sweeps, all nine divergences lay within the range of ideal runs, and so did 135 of the 144 visit frequencies (16 states in each of nine runs). Eight of the remaining nine belong to states with posterior probability below 0.01 
(Fig.~\ref{fig:tasks}e,j,o). Although the measured device signal is correlated in time, the autocorrelation of the state of each p-bit followed the exact curve of ideal Gibbs sampling 
(Fig.~\ref{fig:tasks}d,i,n), so that the correlation of the device noise does not appear as excess correlation in the Markov chain. This includes the slow decay of the competing p-bits in V1, which the exact curve shows as well and which is therefore set by the posterior rather than by the devices. No parameter was fitted to the target distributions.

\subsection*{Inference from new and misleading sensor readings}

\begin{figure*}
\makebox[\textwidth][c]{\includegraphics[width=\textwidth]{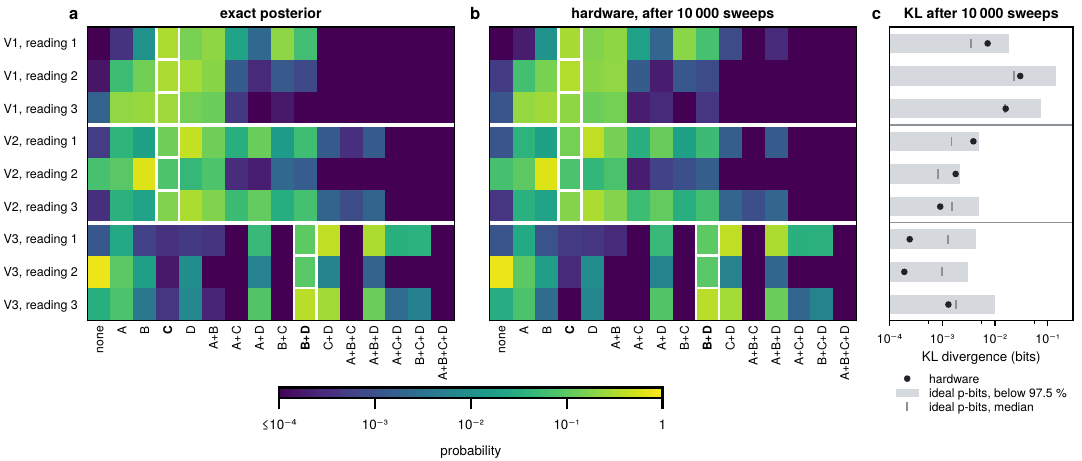}}
\caption{\textbf{Inference from new readings.}
Nine new readings of the  three tasks of Fig.~\ref{fig:tasks}, three per task, each with a new realization of the sensor noise, one row per reading.
\textbf{a}, Exact posterior of the 16 mixtures on a logarithmic color scale, probabilities below $10^{-4}$ in the darkest color. The mixture present (C in V1 and V2, B+D in V3) is framed and its label set in bold.
\textbf{b}, Fraction of the first 10\,000 sweeps of the hardware run spent in each mixture, on the same scale.
\textbf{c}, Kullback--Leibler divergence of these fractions from the posterior (dots), with the range of 1000 simulated runs of ideal p-bits on the same reading below their 97.5\,\% quantile (gray bars) and their median (ticks).}
\label{fig:newdata}
\end{figure*}

The runs above used one reading per task. To test whether the agreement with ideal p-bits holds for data the computer had not seen, we drew nine new readings, three per task, each with a new realization of the sensor noise. Because the couplings depend only on $F$, a new reading changes only the biases $b_i$, so these runs probe the same network at new operating points with unchanged hardware settings. For every reading, the exact posterior and the 95\,\% range of the KL divergence expected from ideal p-bits were computed and recorded before the hardware run, making each run a prediction that the hardware could fail. It did not: all nine divergences
 after 10\,000 sweeps, which span two orders of magnitude from 0.0002 to 0.03 bits depending on the task and reading, lay within the range of ideal p-bits (Fig.~\ref{fig:newdata}c).

These readings also separate inference from optimization. An optimizer, which returns only the most probable state, finds the mixture actually present only when the posterior favors it, and in five of the nine readings the noise points elsewhere. In V3 with the second new reading, the noise makes the absence of all four chemicals the most probable explanation (posterior 0.768) and leaves the mixture actually present, B+D, with only 0.093; the hardware visited these states with frequencies of 0.766 and 0.093. In V2 with the second new reading, chemical B alone becomes the favored explanation (0.636) while chemical C, the one present, falls to 0.076; the hardware returned 0.638 and 0.073 (Fig.~\ref{fig:newdata}a,b). In both cases the computer did not settle on a single answer but reported how strongly the data support each one, including the low but non-zero probability of the true mixture. This is the behavior required of a system that must act on uncertain sensory data, and it is what an optimizer cannot provide.

\subsection*{Autonomous and reprogrammable operation}

The computer also operated autonomously, without supervision or manual intervention. Fifteen of the 18 runs, covering all three tasks with both the original and new readings, were executed consecutively without interruption over about 50\,h. Changing the inference problem required no modification of the hardware. A new sensor model or reading entered solely through the biases and couplings computed by the Python controller, so that competing, degenerate and cooperative posteriors were sampled in succession on the same four p-bits. Each run began with an automated calibration of the p-bits and proceeded to Gibbs sampling, with all parameters set by the controller.

The electrical characteristics of the devices varied appreciably over this period. Between runs, the operating levels of individual memristors shifted by several hundred millivolts, and their noise widths changed by factors of up to six. Nevertheless, every run remained within the range of ideal p-bits, because the trackers continuously updated the baseline, noise width, midpoint and slope of each p-bit. The device in P1, installed before the first run, sampled within the ideal ranges from the outset. The polymer p-bit computer can therefore be reprogrammed entirely in software and operated without supervision, both of which are prerequisites for deployment at the edge.

\subsection*{Comparison with state-of-the-art p-bit hardware}

\begin{figure*}
\makebox[\textwidth][c]{\includegraphics{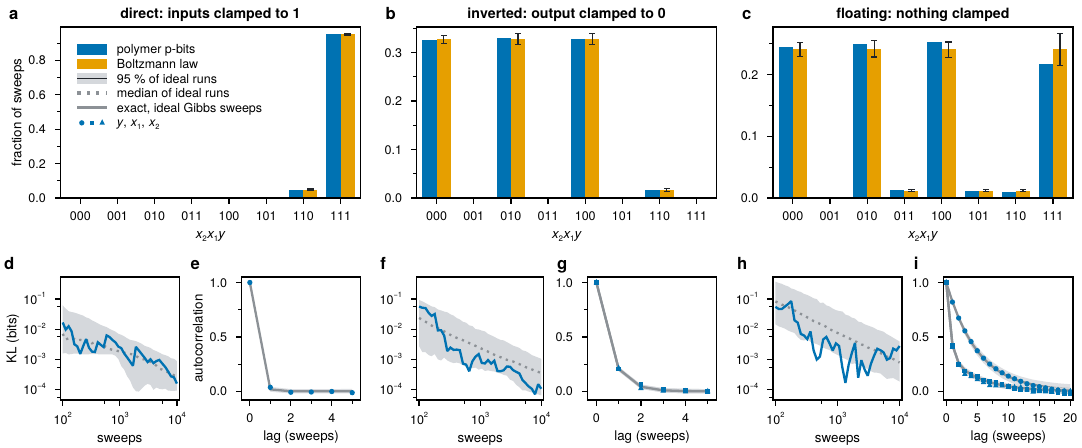}}
\caption{\textbf{The polymer p-bits sample the invertible AND gate of Ref.~\cite{r:borders19} in its three modes.}
The three columns show the direct mode with both inputs clamped to 1 (\textbf{a,d,e}), the inverted mode with the output clamped to 0 (\textbf{b,f,g}) and the floating mode with nothing clamped (\textbf{c,h,i}), each from one run of 10\,000 sweeps with $I_0 = 3$.
\textbf{a--c}, Fraction of sweeps in each of the eight states of the gate, labeled $x_2x_1y$ in the order of Ref.~\cite{r:borders19} (blue), against the Boltzmann law (orange), with the range of 95\,\% of 1000 runs of ideal p-bits on the same schedule (whiskers).
\textbf{d,f,h}, Kullback--Leibler divergence of the gate from its Boltzmann law against the number of sweeps, with the median (dotted) and 95\,\% range (gray) of the ideal runs. 
\textbf{e,g,i}, Autocorrelation of the state of each p-bit that is not clamped, between sweeps ($y$ circles, $x_1$ squares, $x_2$ triangles), that of ideal Gibbs sweeps (line) and the 95\,\% range of the ideal runs (gray).}
\label{fig:and}
\end{figure*}

To compare the polymer computer directly with earlier hardware, we 
also ran the invertible AND gate, the standard benchmark of p-bit networks\cite{r:camsari17prx,r:borders19}. 
Three p-bits represent the inputs $x_1$ and $x_2$ and the output $y$ of the gate, which takes the state 1 only when $x_1 = x_2 = 1$. The energy $E = I_0(3y + x_1x_2 - 2x_1y - 2x_2y)$ of Ref.~\cite{r:borders19}, with $I_0 = 3$, is zero on the four rows of the truth table and positive on the four forbidden states. It was entered as biases and couplings through the Python controller with no change to the circuit design or the calibration procedure. 
Multiplying the energy by $I_0 = 3$ corresponds to p-bits at a three times lower temperature, at which the network moves more slowly between the rows of the truth table and is therefore harder to sample. Run freely, the network spent 96\,\% of sweeps on the four rows of the truth table, in nearly equal shares, as ideal p-bits do (96\,\%). With both inputs clamped to 1, the output took the state 1 in 95\,\% of sweeps, and with the output clamped to 0 the network spent 98\,\% of sweeps on the three consistent input pairs, in both cases as for ideal p-bits. In each mode, the KL divergence to the exact distribution lay within the range of ideal p-bits over the same number of sweeps, and the autocorrelation of each p-bit that was not clamped followed that of ideal Gibbs sampling (Fig.~\ref{fig:and}).
The histograms are not directly comparable with those of the earlier work (ref.~\cite{r:borders19}, Extended Data Fig.~10): there, $I_0$ was adjusted until the Boltzmann law matched the measured counts, whereas here it was set before the runs, and the measured frequencies were compared with the exact distribution without any adjustment.

Table~\ref{tab:compare} compares the polymer computer with the hardware networks of physical p-bits that introduced each active material. In all of them, each p-bit takes the state 1 with a sigmoidal probability of its input, and they differ in the entropy source, in how the input is formed and applied, and in how the result is judged. The magnetic networks use magnetic tunnel junctions patterned by lithography. In the first, a microcontroller computed the input of each p-bit from the states of the others and applied it through a digital-to-analog converter to a transistor in series with the junction\cite{r:borders19}. Later magnetic networks coupled three p-bits directly through a resistor network\cite{r:lv19}, stored learned weights as capacitor voltages while the weighted sum was still formed by a microcontroller\cite{r:kaiser22}, scaled to 80 and 250 junctions driven by a microcontroller or a field-programmable gate array\cite{r:si24,r:yang26}, or integrated the junctions on a CMOS chip whose inputs were computed by a field-programmable gate array\cite{r:li26ne}. In the ferroelectric and oxide-memristor networks, the couplings were summed physically, as currents in a crossbar of binary-weighted ferroelectric junctions\cite{r:guan26} or of multilevel Ru/SiO$_2$ memristors\cite{r:soh26}. A field-programmable gate array or a microcontroller still converted each summed current into the input of a p-bit, and both devices were driven by write pulses for every sample. These systems update far faster than the polymer computer. All addressed optimization, logic or the learning of a fixed mapping, and their success was judged by the most frequent or dominant states\cite{r:borders19,r:guan26,r:soh26}, by a KL divergence from the target distribution without a finite-sampling reference\cite{r:kaiser22}, or by the optimality gap after annealing\cite{r:li26ne}. The polymer computer differs in both material and method. Its active layer is spin-coated from solution and patterned without lithography, and its p-bits fluctuate under a fixed DC bias, the input shifting only the comparator reference. Its chain is run at a constant $\beta$, as Bayesian inference requires, and both its stationary distribution and its mixing are compared with the ranges spanned by ideal p-bits over the same number of sweeps, so that agreement with ideal p-bits is a statistical test against the spread expected from finite sampling rather than a comparison with a single reference.

\begin{sidewaystable}
\caption{Pioneering hardware p-bit networks by active material.}
\label{tab:compare}
\small
\setlength{\tabcolsep}{5pt}
\providecommand{\tc}[2]{\parbox[t]{#1\textwidth}{\raggedright\strut #2\strut}}
\begin{tabular}{@{}lllll@{}}
\toprule
 & \tc{0.20}{\textbf{Magnetic}\\ Borders \textit{et al.}\cite{r:borders19}$^{\ast}$\\ 8 p-bits}
 & \tc{0.20}{\textbf{Ferroelectric}\\ Guan \textit{et al.}\cite{r:guan26}\\ 4 p-bits}
 & \tc{0.20}{\textbf{Oxide memristor}\\ Soh \textit{et al.}\cite{r:soh26}\\ 3 p-bits}
 & \tc{0.21}{\textbf{Polymer}\\ This work\\ 4 p-bits} \\
\midrule
\tc{0.12}{Entropy-source material}
 & \tc{0.20}{CoFeB/MgO stochastic MTJ}
 & \tc{0.20}{Pt/Hf$_{0.5}$Zr$_{0.5}$O$_2$/TiO$_2$/TiN ferroelectric tunnel junction}
 & \tc{0.20}{Pt/SiO$_2$/Ag volatile memristor}
 & \tc{0.20}{ITO/pB6CC/Al conjugated-polymer memristor\cite{r:foukos26acs}}
 \\
\addlinespace
\tc{0.12}{Device fabrication}
 & \tc{0.20}{sputtering, e-beam lithography, ion milling, 300\,$^\circ$C anneal}
 & \tc{0.20}{ALD (320\,$^\circ$C), sputtering, 600\,$^\circ$C RTA}
 & \tc{0.20}{photolithography, e-beam and glancing-angle deposition}
 & \tc{0.20}{spin-coated active layer; shadow-mask Al; no lithography}
 \\
\addlinespace
\tc{0.12}{Validation}
 & \tc{0.20}{agreement with simulation using one fitting parameter}
 & \tc{0.20}{two dominant states match the two maximum independent sets}
 & \tc{0.20}{valid states each $>$20\%}
 & \tc{0.20}{KL within 95\,\% range of ideal p-bits in all 21 runs; autocorrelation as for ideal Gibbs sampling}
 \\
\addlinespace
\tc{0.12}{Task}
 & \tc{0.20}{ground state: integer factorization up to 945; invertible AND}
 & \tc{0.20}{ground state: maximum independent set of a 4-node graph}
 & \tc{0.20}{valid states: reversible AND and NAND}
 & \tc{0.20}{posterior sampling at fixed $\beta$: Bayesian inference (3 tasks, 12 synthetic sensor readings); invertible AND}
 \\
\addlinespace
\tc{0.12}{Coupling implementation}
 & \tc{0.20}{microcontroller + DAC}
 & \tc{0.20}{4$\times$4 FTJ crossbar, binary weights; FPGA}
 & \tc{0.20}{Ru/SiO$_2$ 1R crossbar; digitized by microcontroller}
 & \tc{0.20}{Python controller; one microcontroller per p-bit}
 \\
\bottomrule
\end{tabular}
\par\smallskip
\parbox{0.95\textwidth}{\small\raggedright
$^{\ast}$Other hardware networks of MTJ p-bits: Lv \textit{et al.}\cite{r:lv19} (three p-bits coupled directly through a resistor network, with no controller in the loop; invertible AND gate); Kaiser \textit{et al.}\cite{r:kaiser22} (five p-bits coupled through a microcontroller, weights learned on RC capacitors; full-adder distribution, KL $\approx$0.2 vs target); Si \textit{et al.}\cite{r:si24} (80 superparamagnetic MTJs coupled through a microcontroller and DACs; 9-city traveling-salesman problem, 70-city by graph partitioning); Yang \textit{et al.}\cite{r:yang26} (250 STT-MTJ p-bit cells time-multiplexed across spins by an FPGA; 24-bit factorization, max-cut); Li \textit{et al.}\cite{r:li26ne} (96-kb VC-MRAM on 40\,nm CMOS with FPGA-computed inputs; max-cut, routing with 450 spins and layer assignment).}
\end{sidewaystable}

\section*{Conclusions}

We have shown that a solution-processed conjugated polymer can serve as the entropy source of a working probabilistic computer. Four pB6CC memristors, each with its own resistance, noise level and drift, behave as ideal p-bits once their operating points are tracked, and the resulting network samples Bayesian posteriors with statistics that lie within the finite-sample range of ideal p-bits. This holds for posteriors with competing, degenerate and cooperative structure, for the sampled distribution and for its autocorrelation, for readings whose outcome was predicted before each run, and without any parameter fitted to the target.

The demonstration was designed to obtain the most complete and statistically rigorous answer, and it leaves speed, scale and integration as the natural next steps, none of which is fixed by the architecture. Four p-bits suffice to realize posteriors of qualitatively distinct structure, and larger networks require additional circuits but no change to the controller; a newly installed device in P1 sampled within the ideal ranges from its first run, without conditioning and without changes to the controller. The couplings are computed by a controller, as in most earlier p-bit networks\cite{r:borders19,r:kaiser22,r:li26ne,r:yang26}, and could instead be summed physically by a resistive or capacitive network between the p-bits, as demonstrated for three magnetic p-bits\cite{r:lv19} and partly realized in memristive crossbars\cite{r:guan26,r:soh26}; such clockless networks sample correctly when the coupling is faster than the p-bit fluctuations\cite{r:sutton20}. The update rate is set mainly by the readout and the communication with the controller (Methods). Changes to the microcontroller code alone, made after the runs reported here, have since reduced the time per p-bit update from 0.2 to 0.06\,s with the same sampling statistics, and further gains of this kind are expected without changes to the devices or the circuit. The fluctuation dynamics of the polymer itself can be tuned through its conjugated backbone and pendant groups\cite{r:foulger21jmcc,r:vilkos26ceja}.

Such a computer need not match the speed of silicon. Sensing at the edge unfolds over seconds to minutes, and a probabilistic computer that delivers a calibrated posterior on that timescale beside or within the sensor\cite{r:zhou20} would complement rather than compete with CMOS hardware. Physical p-bits can also be updated in parallel wherever they are not directly coupled\cite{r:aadit22}, so that the time per sweep is set by the fluctuation time of the devices rather than by their number.

Polymer p-bits share the processing advantages that distinguish organic electronics from silicon. The active layer here is spin-coated from solution, and conjugated-polymer memristors have already been fabricated by roll-to-roll gravure printing on flexible ITO\cite{r:grant22amt}, by screen printing combined with vapor-phase polymerization into cascaded logic circuits\cite{r:grant23amt}, and by vapor-phase polymerization over conformal and nonconformal substrates\cite{r:grant23aem}. A printed $N\times N$ array of p-bits joined by printed analog interconnects would perform the coupling summation in the same material layer as the entropy sources, remove the controller from the loop and place the probabilistic computer directly on the surface where data are acquired. Such a computer could infer the state of its environment from noisy sensors, for example the composition of an odor, and report not only its best estimate but also how certain that estimate is.

\section*{Methods}

\subsection*{Polymer synthesis and memristor fabrication}
The memristive layer was pB6CC, a carbazole-based conjugated polyelectrolyte. It was synthesized and characterized as reported previously\cite{r:foukos26acs}. 
Its repeat unit (Supplementary Fig.~\ref{fig:Spolymer}) is a poly(arylene ethynylene) backbone of alternating 3,3$'$-(1,4-phenylenebis(oxy))bis(propane-1-sulfonate) and 3,6-diethynylcarbazole units, with an electrically responsive 9-carbazolyl group tethered to each backbone carbazole nitrogen through a hexyl spacer. The comonomer 9-(6-(9\textit{H}-carbazol-9-yl)hexyl)-3,6-diethynyl-9\textit{H}-carbazole was prepared by \textit{N}-alkylation of 3,6-diethynyl-9\textit{H}-carbazole with 6-(9\textit{H}-carbazol-9-yl)hexyl methanesulfonate (NaH, tetrahydrofuran, reflux). It was then polymerized with sodium 3,3$'$-((2,5-diiodo-1,4-phenylene)bis(oxy))bis(propane-1-sulfonate) by Sonogashira polycondensation (PdCl$_2$(PPh$_3$)$_2$/CuI/PPh$_3$ in dimethylformamide/water/piperidine, 70\,$^\circ$C, 20\,h). The polymer was purified by reprecipitation from dimethylformamide into diethyl ether, giving the piperidinium sulfonate salt. Devices were fabricated on glass substrates (12.7\,mm $\times$ 12.7\,mm $\times$ 0.7\,mm) coated with SiO$_2$-passivated indium tin oxide (ITO, 8--12\,$\Omega$\,sq$^{-1}$). The ITO was patterned into a 4\,mm strip by Zn/HCl etching, cleaned by sonication in acetone and isopropanol (10\,min each), dried under N$_2$ and treated with an oxygen plasma for 5\,min. pB6CC films 200\,nm thick were spin-coated at 2000\,rpm from a 10\,mg\,mL$^{-1}$ solution in 1:1 (v/v) water/tetrahydrofuran. Aluminum top electrodes (nominal thickness 200\,nm, 189\,nm by profilometry) were thermally evaporated at $2\times10^{-6}$\,Torr through a shadow mask orthogonal to the ITO strip. This gave two crossbar ITO/pB6CC/Al memristors per substrate, each with an active area of 4\,mm$^{2}$.

\subsection*{Polymer p-bit circuit}
Each p-bit followed the voltage-divider/comparator architecture we introduced previously for polymer memristors\cite{r:foulger25apr,r:foukos26apr} (Fig.~\ref{fig:unit}a). One ITO/pB6CC/Al memristor formed the upper arm of a divider, with the Al electrode connected to the supply $V_{\mathrm{DD}}$ and the ITO electrode to the divider midpoint. A 1\,M$\Omega$ load resistor formed the lower arm of the divider to ground. Each p-bit was independently biased using a dedicated source-measure unit (SMU) of a Hewlett-Packard 4145A semiconductor parameter analyzer, enabling independent control of the applied supply voltage.
Stochastic conductance fluctuations of the polymer produced corresponding fluctuations in the divider midpoint voltage $V_{\mathrm{mvd}}$, with the voltage across the memristor given by $V_{\mathrm{DD}}-V_{\mathrm{mvd}}$. 

The four primitive circuits were assembled on a solderless breadboard, with the analog signal-conditioning components powered by dedicated supply rails. 
To minimize loading of the voltage divider, $V_{\mathrm{mvd}}$ was isolated from the downstream circuitry using an LF356 JFET-input operational amplifier, 
powered by $\pm$12\,V supply rails and configured as a unity-gain voltage follower. The buffered signal was applied to the inverting input of an LM393 voltage comparator, powered by a single 5\,V supply, while the non-inverting input received a programmable reference voltage $V_{\mathrm{ref}}$, generated by a dedicated 16-bit AD5693R digital-to-analog converter (DAC, Analog Devices). The DAC employed its internal 2.5\,V reference to provide a nominal output range of 0--2.5\,V, corresponding to a voltage resolution of 38.1\,$\mu$V per step. The open-collector of the comparator was connected to a 10\,k$\Omega$ pull-up resistor ($R_l$) tied to the 3.3\,V logic supply of the Teensy 4.1 microcontroller, with a 10\,pF capacitor ($C_l$) to ground for suppression of high-frequency switching transients. The buffered $V_{\mathrm{mvd}}$ and programmed $V_{\mathrm{ref}}$ were independently monitored using two 16-bit LTC1864 successive-approximation analog-to-digital converters (ADCs, Analog Devices), implemented on MikroElektronika ADC 11 Click modules. An SN74AHCT125N quad buffer (Texas Instruments), powered by a 5\,V supply, provided logic-level translation of the 3.3\,V clock and chip-select signals from the microcontroller to the ADCs. The Teensy 4.1 coordinated DAC programming, ADC acquisition and digital comparator readout.
The result is a binary output whose probability of being in state 1 follows a logistic function of the reference offset. Each p-bit's input $\beta x$ is therefore applied through $V_{\mathrm{ref}}$, allowing the output probability to be controlled independently of the memristor supply bias. The four p-bits (P1--P4) were operated at supply voltages of $V_{\mathrm{DD}}=2.0$, 2.0, 1.5 and 2.0\,V respectively.

\subsection*{Implementation of the polymer p-bit computer}

The four p-bits sampled the Boltzmann distribution
\begin{equation}
P(s)=\frac{1}{Z}\,e^{-\beta E(s)},\qquad
E(s)=-\sum_{i}b_is_i-\sum_{i<j}J_{ij}s_is_j,
\label{eq:boltzmann}
\end{equation}
of the joint states $s=(s_1,\dots,s_4)$, where $s_i\in\{0,1\}$ is the
state of p-bit $i$ (1 when $\Vmvd<\Vref$), $b_i$ are the biases,
$J_{ij}=J_{ji}$ are the couplings, $Z$ normalizes the distribution and
$\beta$ is the effective inverse temperature parameter.
Given the states of the other
p-bits, p-bit $i$ is in state 1 with probability
\begin{equation}
P(s_i=1\mid s_j,\,j\neq i)=\frac{1}{1+e^{-\beta x_i}},\qquad
x_i=b_i+\sum_{j\neq i}J_{ij}s_j,
\label{eq:conditional}
\end{equation}
where $x_i$ is the local field of p-bit $i$ and $\beta x_i$ its input.

The p-bits were updated one at a time in the fixed order 1, 2, 3, 4
(Fig.~\ref{fig:unit}b), and one complete 
pass through the four is a sweep.  A Python-based controller running on a Windows PC maintained the stored binary states of the four p-bits and communicated with four Teensy 4.1 microcontrollers, one dedicated to each p-bit, through independent USB serial connections. To update p-bit $i$, the controller
computed $x_i$ from the
stored states of the other three p-bits by Eq.~(\ref{eq:conditional})
and sent the offset $V_{\mathrm{offset},i}$ of Eq.~(\ref{eq:offset})
to the microcontroller of the p-bit, where $\Vc_i$ and $k_i$ are the
midpoint and the slope of its logistic response to the offset (see
Methods section `Probability calibration and tracking').
The microcontroller set
$\Vref=\Vhat+V_{\mathrm{offset},i}$, and the comparator state in the
second reading then
became the new stored state $s_i$. The controller
then updated the next p-bit, on average 0.2\,s after the previous one.
Between its updates a p-bit kept its offset, and its other readings were
used only to track the voltage baseline and the noise width. The
couplings were thus applied digitally through the inputs, and the p-bits
were updated in sequence by the controller, as in earlier experiments
with magnetic tunnel junctions~\cite{r:kaiser22}.
Each update
draws $s_i$ from its conditional distribution under $P$ and leaves $P$
unchanged. For ideal p-bits, the sequence of states is therefore
a Gibbs sampler with a
fixed update order~\cite{r:geman84}, which converges to the Boltzmann
distribution of states from any initial joint state.
Successive joint states are correlated through the couplings. For ideal
p-bits, this correlation follows exactly from the transition
probabilities of one sweep (Supplementary Note~3).

\subsection*{Probability calibration and tracking}

In this section we omit the index $i$ of the p-bit.
The logistic response of each p-bit to the reference offset (see Methods
section `Polymer p-bit circuit') was described
by its midpoint $\Vc$ and slope $k$, which set the offset applied for an
input (see Methods section `Implementation of the polymer p-bit computer'). Both were
calibrated before each run and updated during sampling. We wrote the
slope as $k=\kappa/q$, where $q$ is the noise width of $\Vmvd$ (here and
below, a width is the 10--90\,\% quantile range) and $\kappa$ a
dimensionless scale.  The baseline $\Vhat$ and the noise width $q$
describe the analog signal of the cell (the memristor and its divider)
and were estimated from every reading of the raw voltage.  The midpoint
$\Vc$ and the scale $\kappa$ describe the logistic response of the p-bit
and were estimated from the stored states $s$.

\textit{Voltage baseline $\Vhat$ and noise width $q$.} Each cell was read every
63\,ms. In each reading $n$ its microcontroller first recorded the
comparator state and then measured $\Vmvd(n)$ and $\Vref(n)$.  It
updated the baseline as
\begin{equation}
\Vhat(n)=\Vhat(n-1)+\frac{\Vmvd(n)-\Vhat(n-1)}{\tau_{\mathrm b}},
\label{eq:baseline}
\end{equation}
and held the comparator reference at $\Vref=\Vhat+\Voff$, where
$\Voff$ is the reference offset set by the controller. We set
$\tau_{\mathrm b}=1.5$ readings, near the minimum of the
summed squared autocorrelation of consecutive p-bit states at lags
1--20, obtained by applying Eq.~(\ref{eq:baseline}) to voltages recorded
from a single p-bit at a fixed offset. The
baseline was initialized from five measurements whenever acquisition
started. The controller estimated $q$ as the width of the deviation
$\Vmvd(n)-\Vhat(n-1)$, tracking both quantiles by running estimates
analogous to Eq.~(\ref{eq:baseline}), with $\tau_q=300$ readings.
This value is heuristic and 100 to 3000 readings gave nearly the same
results on recorded runs. The first reading after each change of $\Voff$
was not used for $q$. Every 340\,s, $\VDD$ was set to 0\,V for 30\,s and
restored for 30\,s before sampling resumed.

\textit{Midpoint $\Vc$ and scale $\kappa$ of the logistic response.}
Before each run, the response of each p-bit was measured in two passes.
A coarse pass at 31 fixed offsets gave estimates of $\Vc$ (median) and
of the response width $W$, hence $\kappa=2\ln9\,q/W$, since a logistic
of slope $k$ has the width $W=2\ln9/k$.  The fine pass used 15 offsets
spaced evenly over the range in which these estimates predict 0.1\,\% to
99.9\,\% in state 1, with 400 states each. A maximum-likelihood logistic
fit (Supplementary Note~2) to the fraction in state 1 at these offsets
gave the initial $\Vc$ and $\kappa$.  At each update $m$ of the p-bit,
with input $z(m)=\beta x(m)$, stored state $s(m)$ and predicted
probability $p(m)=1/[1+e^{-z(m)}]$, $\Vc$ and $\kappa$ were updated as
\begin{align}
\Vc(m)&=\Vc(m-1)-\eta_V\,\frac{q}{\kappa(m-1)}\,[s(m)-p(m)],
\label{eq:midpoint}\\
\ln\kappa(m)&=\ln\kappa(m-1)+\eta_\kappa\,z(m)\,[s(m)-p(m)].
\label{eq:slope}
\end{align}
Given the stored states, Eq.~(\ref{eq:midpoint}) moves $\Vc$ toward the
maximum-likelihood estimate of the logistic midpoint, and
Eq.~(\ref{eq:slope}) moves $\kappa$ toward the value at which
$\kappa/q$ is the maximum-likelihood estimate of the slope
(Supplementary Note~2). With a fixed target in place of $p(m)$,
Eq.~(\ref{eq:midpoint}) is the rule used to hold the switching
probability of a magnetic tunnel junction against
drift~\cite{r:zhang25pra}. Both updates are stochastic-gradient steps
with heuristic learning rates $\eta_V=0.052$ and $\eta_\kappa=0.0062$,
which let the midpoint and the scale follow drifts over about 1 and
5\,min of sampling. The extreme tail probabilities of the p-bits may not
follow the logistic law, so a state updated the estimates in
Eqs.~(\ref{eq:midpoint}) and (\ref{eq:slope}) only if $0.01\le
p(m)\le0.99$, i.e., $|z(m)|\le4.6$. Limits between $|z|=3$ and 5 gave
the same results in simulation. Only the midpoint and the slope of the
logistic were calibrated, the empirically measured response was not
remapped, and departures of the devices from the logistic shape remain
in the calibration (Fig.~\ref{fig:unit}e) and in the sampled data.

\subsection*{Inference tasks}

The tasks were Bayesian inference problems in a minimal model of odor
sensing with cross-reactive sensors~\cite{r:persaud82}: each sensor
responds to chemicals A--D, which are all equally likely
to be present, so that which chemicals are present can be inferred only
from the readings of all sensors together. In each task, the sensor readings
$y=(y_1,\dots,y_4)$ were given, and the question was
which of the four chemicals, A--D, were present. With the joint state $s$
describing the chemicals present ($s_i=1$ if chemical $i$ is present),
the answer is the posterior distribution given by Bayes' theorem,
$P(s\mid y)\propto P(y\mid s)\,\pi(s)$, where $\pi(s)$ is the prior.

Sensor $k$ responded to chemical $i$ with $F_{ki}$ (sensor matrix $F$,
Fig.~\ref{fig:tasks}a,f,k), the responses to the chemicals present
added, and each reading had an independent Gaussian error of standard
deviation $\sigma$, i.e.,
$y=Fs^\ast+\sigma\xi$, where $s^\ast$ is the joint
state of the chemicals present and $\xi$ is i.i.d. standard normal. The
likelihood is therefore $P(y\mid s)\propto\exp[-\|y-Fs\|^2/(2\sigma^2)]$.
With the uniform prior, $\pi(s)=1/16$, the logarithm of the
posterior is a quadratic function of $s$, and since $s_i^2=s_i$, the
posterior is the Boltzmann distribution of Eq.~(\ref{eq:boltzmann}) at
$\beta=1$, with
\begin{equation}
b_i=\frac{1}{\sigma^2}\Bigl(\sum_kF_{ki}y_k-\frac12\sum_kF_{ki}^2\Bigr),\qquad
J_{ij}=-\frac{1}{\sigma^2}\sum_kF_{ki}F_{kj}.
\label{eq:posterior}
\end{equation}
The Gibbs sampler (see Methods section `Implementation of the polymer p-bit computer')
therefore has the posterior as its stationary distribution, and the
fraction of sweeps that it spends in a joint state estimates the
posterior probability of that state.

Chemicals A--D were assigned to p-bits 1--4. In V1 (competition, C
present, $\sigma=2$), all couplings are negative and C, B, A+B and D
compete as explanations of the readings. In V2 (symmetry, C present,
$\sigma=3$), the response to C equals the summed responses to A and B,
so that C and A+B have the same posterior probability for any readings.
In V3 (cooperation, B and D present, $\sigma=2.5$), D lowers the readings
of sensors 2 and 4, so that its couplings are positive, and the readings
are explained by B and D together or by neither. The readings were
simulated with the same draw of $\xi$ for the three tasks:
$y=(3.60, 6.69, 1.37, 4.92)$, $(3.90, 7.03, 1.55, 5.39)$ and
$(2.25, -1.14, -1.04, -1.85)$ in V1--V3. Each task was run three times.
Each run started from the joint state $s=(0,0,0,0)$, and we used its first 10\,000
sweeps.

For the AND gate, the output $y$ and the inputs $x_1$ and $x_2$ were assigned to P1, P3 and P4, and P2 was left uncoupled. A p-bit was clamped by adding $\pm(I_0 + \ln 99)$ to its input, the smallest value that keeps it in the clamped state with probability at least 0.99 whatever the states of the other p-bits, and the Boltzmann law includes this term. Calibration and the update order were those of the odor sensing tasks.

\subsection*{Evaluation against ideal p-bits}

After each sweep the fraction $v(s)$ of the joint state $s$ estimated
the posterior probability, whose exact value $P(s)$ was evaluated
directly from Eqs.~(\ref{eq:boltzmann}) and (\ref{eq:posterior}). The
distance of the sampled distribution from
the posterior was measured by the KL divergence
$\kl{v}{P}=\sum_sv(s)\log_2[v(s)/P(s)]$ in bits, summed over the visited
joint states. Fig.~\ref{fig:tasks}b,c shows $v(s)$ and the divergence
every 100 sweeps, and Fig.~\ref{fig:tasks}e shows $v(s)$ after 10\,000
sweeps. The
ranges of 95\,\% in Figs.~\ref{fig:tasks}--\ref{fig:and} were obtained from 1000
simulated runs of ideal p-bits following Eq.~(\ref{eq:conditional}),
each with the same initial state, update order and number of sweeps.
Each range lies between the 2.5 and 97.5\,\% quantiles of the same
quantity over these runs.
The
autocorrelation of the state of each p-bit between sweeps was estimated
from each run and compared with its exact value for ideal
p-bits (Supplementary Note~3).

For each of the nine new readings, drawn with new realizations of $\xi$, the posterior and the 95\,\% range of the divergence after 10\,000 sweeps were computed in the same way and recorded before the hardware run.

The learning rates in Eqs.~(\ref{eq:midpoint}) and (\ref{eq:slope}) are positive numbers, hence the tracked chain of states is not expected to converge exactly to the posterior. The estimates of the midpoint and the slope keep fluctuating around their true values, and therefore the KL divergence from the posterior levels off at a value that decreases with the learning rates. However, in simulations of ideal p-bits with the same online calibration and the learning rates used here, this theoretical KL discrepancy lies two to three orders of magnitude below the KL values reached after the 10\,000 sweeps of a run.
While tracking does not effectively shift the sampled posterior, it reduces the variability of the sampled frequencies\cite{r:hammersley56}. Each update of the midpoint compensates for the discrepancy between the actual answer and its target probability. This explains the apparent ``faster than ideal'' convergence in relevant runs.

\subsection*{Data availability}

The data supporting the findings of this study, including the raw midpoint-voltage readings, comparator states, calibration data, controller logs and run reports for all hardware runs presented here, are available in Zenodo at \url{https://doi.org/10.5281/zenodo.XXXXXXX}. Source data for Figs.~1--4 are provided with this paper.

\subsection*{Code availability}

The controller code used to operate the polymer p-bit computer and the code used to compute the posteriors, simulate ideal p-bits and analyze the runs are available in Zenodo at \url{https://doi.org/10.5281/zenodo.XXXXXXX}.

\clearpage
\section*{Supplementary Information}
\setcounter{equation}{0}
\setcounter{figure}{0}
\renewcommand{\thefigure}{S\arabic{figure}}
\renewcommand{\theequation}{S\arabic{equation}}

\subsection*{Supplementary Note 1. Polymer and memristor}

\begin{figure}[h]
\centering
\includegraphics[scale=0.5]{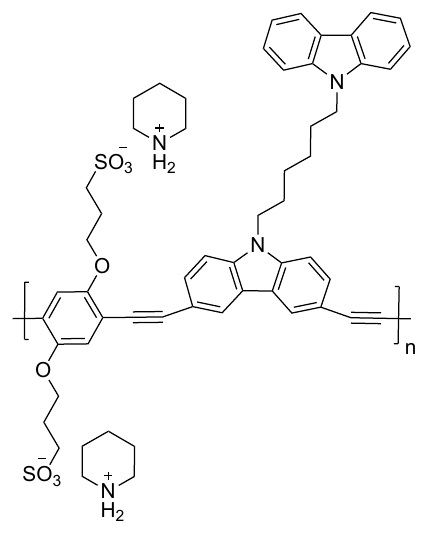}
\caption{{\textbf{Polymer repeat unit.} Repeat unit of pB6CC: alternating 2,5-bis(3-sulfonatopropoxy)-1,4-phenylene and 9-[6-(9\textit{H}-carbazol-9-yl)hexyl]-9\textit{H}-carbazole-3,6-diyl units joined by ethynylene links, with piperidinium counterions\cite{r:foukos26acs}.}}
\label{fig:Spolymer}
\end{figure}

The memristive layer is pB6CC\cite{r:foukos26acs}, a conjugated polyelectrolyte with a poly(arylene ethynylene) backbone of alternating 3,3$'$-(1,4-phenylenebis(oxy))bis(propane-1-sulfonate) and 3,6-diethynylcarbazole units and a 9-carbazolyl pendant group tethered to each backbone carbazole nitrogen through a hexyl spacer (Supplementary Fig.~\ref{fig:Spolymer}). Each memristor is a crossbar ITO/pB6CC/Al sandwich with a 200\,nm polymer film and an active area of 4\,mm$^2$ (Methods).

\subsection*{Supplementary Note 2. Maximum-likelihood calibration and tracking of the p-bit response}

The midpoint and the scale of each p-bit were initially fitted during the calibration period and then updated after every state update as follows. 

\textit{Log-likelihood of a stored state.} With the probability $p=1/(1+e^{-z})$ of state 1 at the
input $z=k(\Voff-\Vc)$, $k=\kappa/q$ (Methods), a stored state $s\in\{0,1\}$ has the log-likelihood
$\ln L=s\ln p+(1-s)\ln(1-p)$. Since $\mathrm{d}p/\mathrm{d}z=p(1-p)$, its derivatives with respect to
the midpoint in units of the input, $k\Vc$, and to $\ln\kappa$ are,
\begin{equation}
\frac{\partial\ln L}{\partial(k\Vc)}=-(s-p),\qquad
\frac{\partial\ln L}{\partial\ln\kappa}=z\,(s-p).
\label{eq:Sscore}
\end{equation}

\textit{Fit before the run.} For the states $s(n)$, $n=1,\dots,N$, of the fine calibration pass, the
maximum-likelihood estimates of $\Vc$ and $\kappa$ make both derivatives vanish when summed over the
states, $\sum_n[s(n)-p(n)]=0$ and $\sum_n z(n)\,[s(n)-p(n)]=0$. The solution is unique and was found by Newton
iteration.

\textit{Update after each stored state.} During sampling the response drifts. Instead of repeating the
fit, each stored state moves the estimates a small step along the derivatives of
Supplementary Eq.~(\ref{eq:Sscore}),
\begin{equation}
k\,\Delta\Vc=-\eta_V\,(s-p),\qquad \Delta\ln\kappa=\eta_\kappa\,z\,(s-p),
\label{eq:Sstep}
\end{equation}
which, with $1/k=q/\kappa$, are Eqs.~(\ref{eq:midpoint}) and (\ref{eq:slope}) of the Methods. These are
stochastic-gradient steps on the log-likelihood with constant learning rates. On average they vanish
where the two sums of the fit vanish, so that $\Vc$ and $\kappa$ follow the maximum-likelihood estimates
for the recent states. The learning rates are heuristic. Near $z=0$ an error in $\Vc$ decays over
$4/\eta_V=77$ updates, and near $|z|=2.4$, where a state carries the most information on the slope, an
error in $\ln\kappa$ decays over about 370 updates, that is, over about 1 and 5\,min of sampling. 

\subsection*{Supplementary Note 3. Correlation between sweeps}

\textit{Transition probabilities.} The update of p-bit $i$ takes the
joint state $s=(s_1,\dots,s_4)$ to the joint state $s'$ with
probability~\cite{r:geman84}
\begin{equation}
T_i(s'\mid s)=
\begin{cases}
P(s'_i\mid s_j,\,j\neq i), & s'_j=s_j \text{ for all } j\neq i,\\
0, & \text{otherwise},
\end{cases}
\label{eq:STi}
\end{equation}
where $P(s'_i\mid s_j,\,j\neq i)$ is given by Eq.~(\ref{eq:conditional})
of the Methods for $s'_i=1$ and is one minus that value for $s'_i=0$.

\textit{Exact autocorrelation.} Let $T^{(\ell)}(s'\mid s)$ be the
transition probability from state $s$ to $s'$ over $\ell$ sweeps, each
sweep updating p-bits 1 to 4 in turn. For a chain in its stationary
distribution, the Boltzmann distribution $P$ of Eq.~(\ref{eq:boltzmann}) of the Methods,
the autocorrelation of the state of p-bit $i$ at lag $\ell$
is~\cite{r:sokal95}
\begin{equation}
\rho_i(\ell)=\frac{\sum_{s,s'}P(s)\,s_i\,T^{(\ell)}(s'\mid s)\,s'_i-m_i^2}{m_i(1-m_i)},
\qquad m_i=\sum_s P(s)\,s_i,
\label{eq:Sacf}
\end{equation}
which gives the lines in Fig.~\ref{fig:tasks}d,i,n.

\textit{Estimate from a run.} With $x_t$ the state of p-bit $i$ after
sweep $t=1,\dots,n$ ($n=10\,000$) and $\bar x$ their mean,
$\rho_i(\ell)$ was estimated as
\begin{equation}
\hat\rho_i(\ell)=\frac{\sum_{t=1}^{n-\ell}(x_t-\bar x)(x_{t+\ell}-\bar x)}{\sum_{t=1}^{n}(x_t-\bar x)^2}.
\label{eq:Sest}
\end{equation}
Its distribution at this length has no closed form for a binary chain.
The bands in Fig.~\ref{fig:tasks}d,i,n are therefore the 2.5 and
97.5\,\% quantiles of $\hat\rho_i(\ell)$ over 1000 simulated runs of
ideal p-bits with the same 10\,000 sweeps (Methods), which include the
bias and the spread of the estimator.

\linespread{0.97}
\bibliography{references}

\section*{Acknowledgements}

This work was supported by the Gregg-Graniteville Foundation, the Czech Science Foundation (project No.~24-10384S) and the Ministry of Education, Youth and Sports of the Czech Republic (MEYS; project Inter-Excellence No.~LUAUS26281).

\section*{Author contributions}

Y.B. synthesized the polymer. T.W. fabricated the devices, built the probabilistic computer, wrote its control code and operated it. L.K. developed the inference tasks and sampling algorithms and implemented them on the computer. S.H.F. conceived the polymer p-bits and directed all phases of the construction and testing of the computer. All authors wrote the manuscript.

\section*{Competing interests}

S.H.F., T.W. and Y.B. are inventors on a provisional patent application related to this work, filed by the Clemson University Research Foundation. The authors declare no other competing interests.

\end{document}